\documentclass[aps,prl,twocolumn,graphicx,showpacs]{revtex4-1}
\usepackage{amsmath,amssymb,amsfonts}
\usepackage{bbold}
\usepackage{color}
\usepackage{url}
\usepackage{graphicx}

\draft 

\newcommand{\leb}[1]{\text{d}[#1]}
\newcommand{\dd}[1]{\text{d}#1}
\newcommand{\Det}{\text{det}}

\newcommand{\tr}{\text{tr}}
\newcommand{\Id}{\mathbb{1}}

\newcommand{\diag}{\text{diag}}

\newcommand{\dagg}{^{\dagger}}

\begin{document}

\title{Distribution of the Smallest Eigenvalue in the Correlated Wishart Model}

\author{Tim Wirtz}
\email[]{tim.wirtz@uni-due.de}
\author{Thomas Guhr}
\email[]{thomas.guhr@uni-due.de}
\affiliation{Fakult\"at f\"ur Physik, Universit\"at Duisburg--Essen, 47048 Duisburg, Germany}

\date{\today}

\begin{abstract}
  Wishart random matrix theory is of major importance for the analysis
  of correlated time series. The distribution of the smallest
  eigenvalue for Wishart correlation matrices is particularly
  interesting in many applications. In the complex and in the real
  case, we calculate it exactly for arbitrary empirical eigenvalues,
  \textit{i.e.}, for fully correlated Gaussian Wishart ensembles. To
  this end, we derive certain dualities of matrix models in ordinary
  space. We thereby completely avoid the otherwise unsurmountable
  problem of computing a highly non--trivial group integral. Our
  results are compact and much easier to handle than previous ones.
  Furthermore, we obtain a new universality for the distribution of
  the smallest eigenvalue on the proper local scale.
\end{abstract}


\maketitle 

In a large number of complex systems, time series are measured which
yield rich information about the dynamics but also about the
correlations.  Examples are found in physics, climate research,
biology, medicine, wireless communication, finance and many other
fields~\cite{chatfield,Kanasewich,TulinoVerdu,Gnanadesikan,BarnettLewis,VinayakPandey,AbeSuzuki,Muelleretal,Seba,SanthanamPatra,LalouxCizeauBouchaudPotters,ple02}.
Consider a set of $p$ time series $M_j$, $\ j=1,\ldots,p$ of $n$ ($n\geq p$) time steps each, which are normalized to zero mean and unit variance . The entries $M_j(t)$, $t=1,\dots,n$, are either real or
complex, these two cases are labeled by $\beta=1$ or $\beta=2$,
respectively. The $p$ time series form the rows of the rectangular
$p\times n$ data matrix $M$. The empirical correlation matrix of these
data,
\begin{align} 
C = \frac{1}{n} MM\dagg \ ,
\label{definitionCmatrix}
\end{align}
is positive definite and either real symmetric or Hermitian for
$\beta=1,2$.  Wishart random matrix theory plays a prominent role for
the study of statistical features  \cite{muirhead,VinayakPandey,Johnstone,Muelleretal,AbeSuzuki,Seba,SanthanamPatra,TulinoVerdu,LalouxCizeauBouchaudPotters,ple02,ForresterHughes,ChenTseKahnValenzuela}.
The ensemble of Wishart correlation matrices $WW\dagg/n$ is
constructed from $p\times n$ random matrices $W$ such that it fluctuates
around the empirical correlation matrix $C$.  The probability
distribution for this ensemble is usually chosen as the Gaussian
\cite{muirhead}
\begin{align}
P_\beta(W|C) \sim \exp\left(-\frac{\beta}{2}\tr~WW\dagg C^{-1}\right) \ ,
\label{WRMdistribution}
\end{align}
where the dagger simply indicates the transpose for $\beta=1$.  The
corresponding volume element or measure $\leb{W}$ and all other
measures $\leb{\cdot}$ occurring later on are flat, \textit{i.e.},
they are the products of the independent differentials. The Wishart
correlation matrices $WW\dagg/n$ yield upon average the empirical
correlation matrix $ C$.  Invariant observables depend only on the
always non--negative eigenvalues $\Lambda_j,$  $ j=1,\ldots,p$, of $C$
which are referred to as the empirical ones.  We order them in the
diagonal matrix $\Lambda$.  Data analyses strongly corroborate
the Gaussian Wishart model, see \textit{e.g.} Refs.~\cite{TulinoVerdu,AbeSuzuki,Muelleretal,Seba,SanthanamPatra,LalouxCizeauBouchaudPotters}.

The smallest eigenvalue of the Wishart correlation matrix 
$WW\dagg/n$, or equivalently, of $WW\dagg$ is of considerable interest for statistical analysis, from a general
viewpoint and in many concrete applications.  In linear discriminant
analysis it gives the leading contribution for the \textit{threshold estimate}
\cite{LarryWasserman}. It is most sensitive to \textit{noise} in the
data \cite{Gnanadesikan}.  In linear principal component analysis, the
smallest eigenvalue determines the \textit{plane of closest fit}
\cite{Gnanadesikan}.  It is also crucial for the identification of
\textit{single statistical outliers} \cite{BarnettLewis}. In numerical
studies involving large random matrices, the \textit{conditional
  number} is used, which depends on the smallest eigenvalue \cite{Edelman1992,Edelmann1988}.  In wireless communication $W$ models the Multi--Input--Multi--Output
(MIMO) channel matrix of an antenna system \cite{FoschiniGans}.  The
smallest eigenvalue of $C$ yields an estimate for the \textit{error of
  a received signal}
\cite{Burel02statisticalanalysis,ChenTseKahnValenzuela,UpamanyuMadhow}.
In finance, the \textit{optimal portfolio} is associated with the
eigenvector to the smallest eigenvalue of the covariance matrix, which
is directly related to the correlation matrix \cite{Markowitz}.  This
incomplete list shows the considerable theoretical and practical
relevance \cite{muirhead,Johnstone} to study the distribution
$\mathcal{P}^{(\beta)}_{\footnotesize \text{min}}(t)$ of the smallest
eigenvalue. For given empirical eigenvalues $\Lambda_j, \
j=1,\ldots,p$, one has~\cite{muirhead,MehtasBook}
\begin{align}
  \mathcal{P}^{(\beta)}_{\footnotesize\text{min}}(t) = - \frac{\text{d}}{\text{d} t}E^{(\beta)}_{p}(t) \ , 
\label{pminE}
\end{align}
where $E^{(\beta)}_{p}(t)$ is the gap probability that all eigenvalues of $WW^{\dagger}$ lie in $[t,\infty)$. 

We have three goals: First, we calculate the above quantities exactly.
In the real case, we provide, for the first time, explicit and
easy--to--use formulas for applications.  Second, we uncover mutual
dualities between matrix models which make the calculation possible.
Third, we find a new universality on a local scale referred to as
microscopic in Chiral Random Matrix Theory~\cite{Shuryak1993306,VerbaarschotWettig}.

To begin with, we diagonalize $WW\dagg= V X V \dagg$ with $V\in
\text{U}(p)$ if $\beta=2$ or $V\in \text{O}(p)$ if $\beta=1$. The
eigenvalues are non--negative and ordered in the diagonal matrix
$X=\diag(x_1,\dots,x_p)$.  The volume element transforms as
\begin{align}
 \leb{W} &= \left|\Delta_p(X)\right|^{\beta}\Det^{\gamma}X \, \leb{X}\dd{\mu( V)} \ ,
\label{volumeelement}
\end{align}
where $\dd{\mu}(V)$ is the Haar measure and $\Delta_p(X)$ is the
Vandermonde determinant \cite{MehtasBook}. We introduce
\begin{eqnarray}
\gamma &=& \frac{\beta}{2}(n-p+1) -1 
                = \left\{\begin{array}{ll} (n -p -1 )/2, & \ \beta=1 \\ 
                                                                n-p, & \ \beta=2
                             \end{array}\right. ,
\label{gamma}
\end{eqnarray}
which involves the ``rectangularity'' $n-p$ of the matrix $W$.  Thus,
the joint distribution of the eigenvalues reads
\begin{eqnarray}
P_\beta(X|\Lambda) &=K_{p\times n} \left|\Delta_p(X)\right|^{\beta} \Det^{\gamma}X 
              \, \Phi_\beta(X,\Lambda^{-1}) \ ,
\label{jpdfX}
\end{eqnarray}
with the normalization constant $K_{p\times n}$. The highly
non--trivial part is the group integral
\begin{align}
  \Phi_\beta(X,\Lambda^{-1}) &= \int\dd{\mu( V)}\exp\left(-\frac{\beta}{2} 
   \tr V X V\dagg \Lambda^{-1}\right) \ .
\label{gi}
\end{align}
The gap probability can then be cast into the form \cite{MehtasBook}
\begin{align}
\begin{split}
E^{(\beta)}_{p}(t) &= K_{p\times n} \exp\left(-\tr\frac{\beta t}{2\Lambda}\right)
                                       \int\leb{X} \left|\Delta_p(X)\right|^{\beta}\\
                                &\times\Det^{\gamma}\left(X+t\Id_p\right) \, \Phi_\beta(X,\Lambda^{-1}) \ ,
\end{split}
\label{definitionE}
\end{align}
where $\Id_p$ is the $p\times p$ dimensional unit matrix. Importantly,
the derivation of Eq.~\eqref{definitionE} involves the shift $X \to
X+t\Id_p$.  The group integral in \eqref{gi} is known exactly for
$\beta=2$, it is the Harish--Chandra--Itzykson--Zuber integral
\cite{ItzyksonZuber,HarishChandra}. For $\beta=1$, it is the orthogonal
Gelfand spherical function \cite{GelfandSpherical} or the orthogonal
Itzykson--Zuber integral.  Unfortunately, explicit results are not
available, although the real case $\beta=1$ is the much more relevant
one in applications. To make progress, zonal or Jack polynomials
\cite{muirhead} were developed, which are only given by complicated
recursions. The resulting formulas for observables are therefore
cumbersome. In Refs.~\cite{RecheretalPRL,Recheretal}, when calculating
the spectral density, we circumvented this severe problem by employing
the Supersymmetry method \cite{efetov,Verbaarschotetsal}.  In the
present context, a supermatrix model is quickly constructed by
multiplying the right--hand side of Eq.~\eqref{definitionE} with
$\Det^{\gamma}X/\Det^{\gamma}X=1$ which completes the Jacobian
according to Eq.~\eqref{volumeelement}. This allows one to reintroduce
the matrices $W$. The remaining determinants
$\Det^{\gamma}\left(WW\dagg+t\Id_p\right)$ and $\Det^{\gamma}WW\dagg$ in the
numerator and the denominator, respectively, then inevitably lead to a
\textit{supermatrix} model.

Here, we put forward a different approach which will eventually lead
us to a much more convenient matrix model in \textit{ordinary} space.  Anticommuting
variables will only be used in intermediate steps. Our key idea is to
identify rectangular matrices $\overline{W}$ of dimension
$p\times\bar{n}$, with $\bar{n}$ yet to be determined such that the gap
probability acquires the form of a matrix model in $\overline{W}$
\textit{without} a determinant in the denominator. As one sees from
Eq.~\eqref{volumeelement}, this is achieved, if $\Det^{\beta(\bar{n}-p
  +1)/2- 1}X$ becomes unity, \textit{i.e.}, if the condition
\begin{equation}
\bar{n}=p+\frac{2}{\beta}-1=p+2-\beta \ , \quad \textrm{for} \  \beta=1,2 \ ,
\label{condition}
\end{equation}
is fulfilled.  We thus arrive at the matrix model 
  \begin{align}
    \begin{split}
      E^{(\beta)}_{p}(t) &= K_{p\times \bar n} \exp\left(-\tr
        \frac{\beta t}{2\Lambda}\right) \\&\times\int\leb{\overline W}
      \Det^{\gamma}(\overline {W}~\overline
      {W}\dagg+t\Id_{p})\\&\times\exp\left(-\frac{\beta}{2} \tr\overline
        W~\overline W\dagg\Lambda^{-1}\right)~,
    \end{split}\label{finalE}
\end{align}
which is dual to the model~\eqref{definitionE}. Since the only
determinant is in the numerator, we just need anticommuting
variables to lift $\overline {W}~\overline{W}\dagg$ in the exponent
and can carry out the ensemble average. Along lines similar to,
\textit{e.g.}, Refs.~\cite{GuhrWettigWilke,RecheretalPRL,Recheretal},
we then find
\begin{align}
  \begin{split}
  E^{(\beta)}_{p}(t)&=K_{p\times\bar n} 
               \exp\left(-\tr\frac{\beta t}{2\Lambda}\right) \int\leb{\sigma} \exp\left(-\tr \sigma \right) \\ 
                               &\times f_{\beta,\bar n}(\sigma)\prod_{k=1}^p\Det^{\beta/2}
                \left(\frac{\beta t}{2}\Id_{ 2\gamma/\beta}-\Lambda_k\sigma\right) \ ,
\end{split}
\label{finalESUSY}
\end{align}
where we restrict ourselves to integer $\gamma$ for $\beta=1$.
Although we used anticommuting variables, the
$2\gamma/\beta\times2\gamma/\beta$ matrix $\sigma$ is
ordinary~\cite{GuhrWettigWilke}, it is either an Hermitian ($\beta=2$)
or a self dual Hermitian matrix ($\beta=1$). The function
\begin{align}
f_{\beta,\bar n}(\sigma)=\int\leb{\varrho}\Det^{\beta \bar n /2}\varrho \,
                \exp\left(-\imath \tr\varrho\sigma\right)
\label{Ingham-SiegelFF}
\end{align}
is related to the Ingham--Siegel integral, see Ref.~\cite{Fyodorov}.  It is
obviously invariant, $f_{\beta,\bar n}(\sigma)=f_{\beta,\bar n}(s)$, where
$\sigma=usu^{-1}$ with $s$ being the diagonal matrix of the eigenvalues
and where $u \in \text{USp}(\gamma)$ for $\beta=1$ and $u \in
\text{U}(\gamma)$ for $\beta=2$. Using
Refs.~\cite{Guhr2006I,KieburgGroenqvistGuhr}, we conclude that
\begin{align}
 f_{\beta,\bar n}(s) & \sim \prod_{i=1}^{\gamma}
                \frac{\partial^{\bar n+2(\gamma-1)/\beta}}
                         {\partial s_i^{\bar n+2(\gamma-1)/\beta}} \delta(s_i) \ .
\label{InghamSiegelDistResult}
\end{align} 
Remarkably, the ordinary matrix model~\eqref{finalESUSY} is invariant,
once  $f_{\beta,\bar n}(s)$ is evaluated. There is no symmetry breaking which
would lead to an Itzykson--Zuber--type--of integral as is present in
the above mentioned \textit{supermatrix} model. Since the latter
involves for $\beta=1$ an explicitly unknown supergroup integral, the
model~\eqref{finalESUSY} is much better tractable. By constructing the
model~\eqref{finalESUSY} in \textit{ordinary space}, we fully
outmaneuvered the substantial difficulties related to the orthogonal
Itzykson--Zuber integral and to the zonal or Jack polynomials.  Due to
the lack of Efetov--Wegner terms in ordinary space, even the complex
case $\beta=2$ is considerably easier to treat. We mention in passing
that for $\beta=1$ an half--integer value of $\gamma$ enforces a
supermatrix model, but this will be discussed
elsewhere~\cite{WirtzGuhrII}.  

Hence, applying standard techniques, we arrive at
\begin{align}
      E^{(\beta)}_{p}(t) &= \frac{\displaystyle 
                        \exp\left(-\tr\frac{\beta t}{2\Lambda}\right)}{\Det^{\gamma}\Lambda}
                          \Det^{\beta/2}\left[Q^{(\beta,p)}_{ij}(t)\right] \ ,
\label{FinalExactEbeta1}
\end{align}
where the elements of this and all other determinants run over $i,j$
with $i,j=1,\ldots,2\gamma/\beta$. The kernel in
Eq.~(\ref{FinalExactEbeta1}) is a finite polynomial in $t$,
\begin{align}
Q^{(\beta,p)}_{ij}(t) &= q_{ij} \, \Theta(\alpha_{p,\beta})
                                     \sum^{\text{min}(p,\alpha_{p,\beta})}_{k=0}
                                 \frac{ e_k(\Lambda)~t^{p-k}}{(\alpha_{p,\beta}-k)!} \ .
 \label{detkernelbeta1}
\end{align}
Here we defined $q_{ij}=(j-i)(-1)^{j+i}$ and $q_{ij}=(-1)^{i+1}$ for
$\beta=1,2$, respectively, as well as
$\alpha_{p,\beta}=p+2(\gamma+1)/\beta -i-j$. We also introduced the
Heaviside step function $\Theta(x)$ and the elementary symmetric
polynomials
\begin{eqnarray}
e_k(\Lambda) &= \sum\limits_{1\leq i_1 <\dots< i_k \leq p} 
                            \Lambda_{i_1}\cdots\Lambda_{i_k} \ ,
\label{symfct}
\end{eqnarray}
with $k=0,\ldots,p$. Applying Eq.~(\ref{pminE}), we obtain the 
distribution of the smallest eigenvalue in the explicit form 
\begin{align}
     \begin{split}
       \mathcal{P}^{(\beta)}_{\footnotesize \text{min}}(t) &=
       \tr\frac{\beta}{2\Lambda}E^{(\beta)}_{p}(t) -
       \frac{\beta}{2}\frac{\displaystyle\exp\left(-\tr\frac{\beta
             t}{2\Lambda}\right)}{\Det^{\gamma}\Lambda}\\&\times\sum\limits_{l=1}^{2\gamma/\beta}\frac{\Det
         \left[G^{(l)}_{ij}(t) \right]}{\Det^{1-\beta/2}\left[Q^{(\beta,p)}_{ij}(t)\right]}\ ,
     \end{split}  
\label{pminexactbeta1}
\end{align}
where another polynomial kernel occurs,
\begin{align}
G^{(l)}_{ij}(t) &=\left\{ \begin{array}{ll}\displaystyle Q^{(\beta,p)}_{ij}(t) &, l\neq i \\ 
                             \displaystyle\frac{\text{d}}{\text{d} t} Q^{(\beta,p)}_{ij}(t) &, l= i 
                                        \end{array}\right. \ .
\label{gdef}
\end{align}
These results are exact and valid for all integer values of $\gamma$.
As already mentioned, half--integer values are possible for $\beta=1$.

\begin{figure}[htb]
  \centering
  \includegraphics[width=0.48\textwidth]{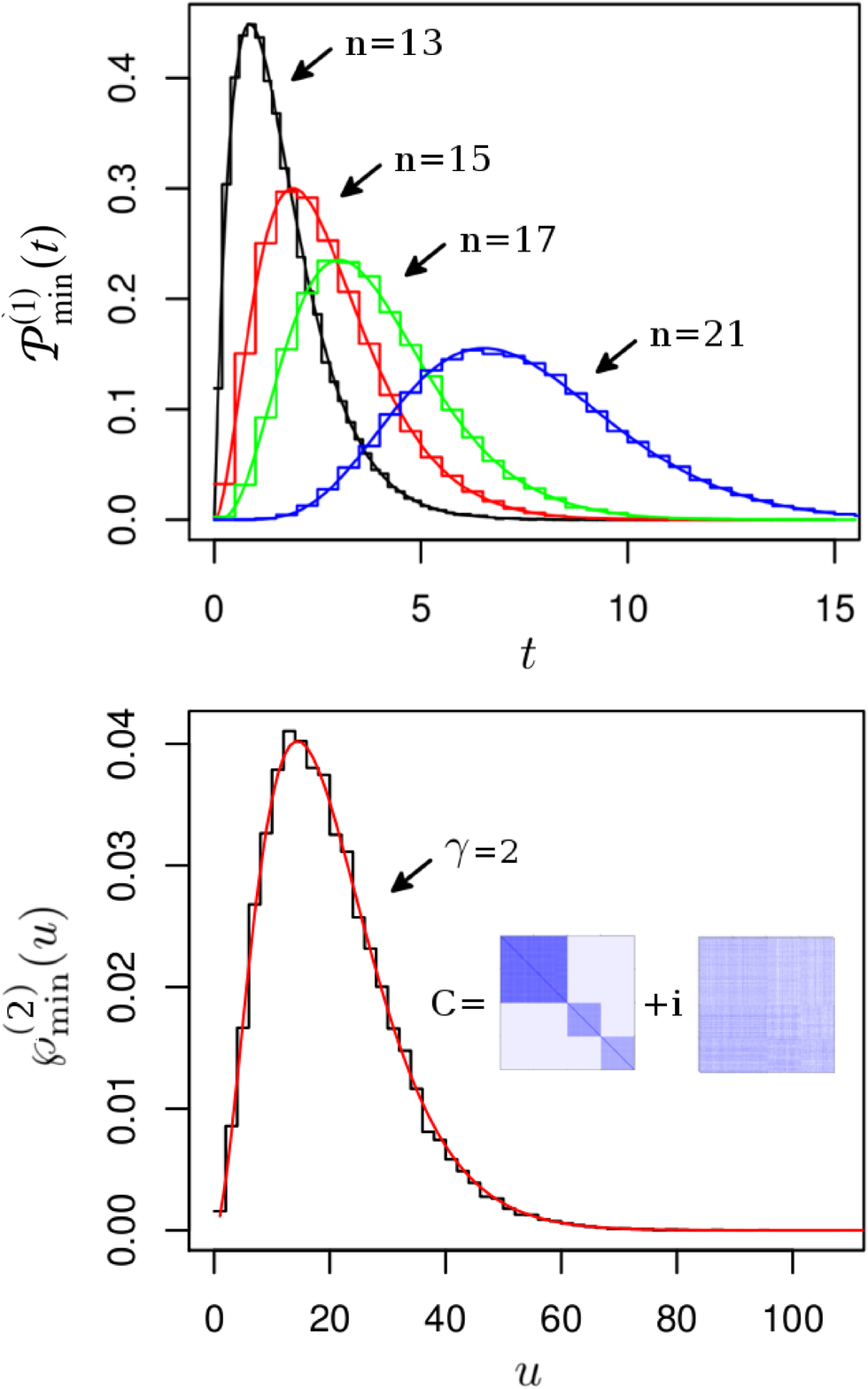}
  \caption{(color online) Distribution of the smallest eigenvalue (top) $\mathcal P^{(1)}_{\footnotesize \text{min}}(t)$ for finite $p=10$, $n=13,15,17,21$ and $\beta=1$,  and (bottom) $\wp^{(2)}_{\footnotesize\text{min}}(u)$   for $\gamma=2$ and $\beta=2$ in the case of correlated Wishart ensembles.  The solid lines correspond to analytic results and the histograms to the numerical simulations.}
  \label{fig}
\end{figure}

Our formula are much more compact and also easier to handle than the
previously known expressions. The duality which we uncovered leads to
much clearer structures. In Ref.~\cite{koev}, $E^{(\beta)}_{p}(t)$ and
$\mathcal{P}^{(\beta)}_{\footnotesize \text{min}}(t)$ for
$\beta=1,2,4$ are expressed, apart from an exponential, in terms of a
finite series in zonal polynomials.  Unfortunately, the latter are
only given recursively and are thus cumbersome in applications. Even
for $\beta=2$, where the Itzykson--Zuber integral is explicitly known,
our formulas have a more direct structure and are more convenient in
applications than the ones in
Refs.~\cite{ForresterminEig,ZhangNuiYangZahngYang}.  We illustrate our
findings in Fig.~\ref{fig}  for $p=10$ real
time series ($\beta=1$) of lengths $n=13,15,17,21$. As empirical
eigenvalues $\Lambda_j$, we chose 0.6, 1.2, 6.7, 9.3, 10.5, 15.5, 17.2,
20.25, 30.1, 35.4.  To demonstrate the validity of our results, we
compare them with numerical simulations. Using the program R \cite{RMan},
we generate and diagonalize 50,000 correlated random Wishart matrices
drawn from the distribution (\ref{WRMdistribution}). As expected, the
agreement is perfect. In contrast to the formulae in Ref.~\cite{koev},
numerical evaluation of Eq.~\eqref{pminexactbeta1} is easily
possible even for large matrix dimension $p$.

We now show the existence of a new kind of universality in the
correlated Wishart model. In addition to the theoretical interest in
this issue, the universal results to be presented now might also be of
practical importance when analyzing correlation matrices built from
many long time series. We want to zoom into a local scale set by the
mean level spacing of the spectral density near the origin. To this
end, we consider the limit $n,p \longrightarrow\infty$ in such a way
that the ``rectangularity'' $n-p$ is held fixed. A similar limit,
referred to as \textit{microscopic}, was introduced in Chiral Random
Matrix Theory \cite{Shuryak1993306,VerbaarschotWettig}, which models statistical aspects of Quantum Chromodynamics. However, in contrast to our case the probability density is fully
rotation invariant, corresponding to  $\Lambda=\Id_p$.  Here we consider an
arbitrary $\Lambda$, and thus have to take into account how the local
mean level spacing depends on $\Lambda$.  From
Refs.~\cite{Forrester1993709,ZhangNuiYangZahngYang} it follows that we
have to rescale $t$ with $p$, when taking the limit $p
\longrightarrow\infty$. Hence we introduce a new variable $u$ by
making the ansatz
\begin{eqnarray}
t = \frac{u}{4 p\eta}
\label{umicro}
\end{eqnarray}
with a $\Lambda$ dependent constant $\eta$ yet to be determined. We thus
we restrict our analysis to the commonly encountered situation in
which almost all empirical eigenvalues do not depend on $p$, and only
few are proportional to $p$. In the microscopic limit, the gap
probability and the distribution of the smallest eigenvalue are
defined as
\begin{align}
    \mathcal{E}^{(\beta)}(u) &= \lim_{p\rightarrow\infty} E^{(\beta)}_{p}\left(\frac{u}{4p\eta}\right)~,\\
    \wp_{\footnotesize\text{min}}^{(\beta)}(u) &= \lim_{p\rightarrow
      \infty} \frac{1}{4p\eta}
    \mathcal{P}_{\footnotesize\text{min}}^{(\beta)}\left(\frac{u}{4 p\eta}\right)
 \ .
\label{microdefs}
\end{align}
The limit $p\rightarrow \infty$ is non--trivial, because the function
$f_{\beta,\bar n}(\sigma)$  and the normalization constant depend on $p$. The
$p$ dependence of the latter is determined by evaluating
Eq.~\eqref{finalESUSY} at $t=0$, which shows that $K_{p\times \bar n} =
(-1)^{\gamma p}\Det^{-\gamma}\Lambda~K_{\gamma}$ with $K_\gamma$
independent of $p$. Next we investigate how the $p$ fold product of
determinants in Eq.~\eqref{finalESUSY} behaves on the local scale,
\begin{align}
  \begin{split}
    &\prod_{k=0}^p\Det^{\beta/2}\left(\frac{\beta u}{8p\eta}\Id_{
        2\gamma/\beta}-\Lambda_k\sigma\right) = 
    \Det^{\gamma}\Lambda \Det^{p\beta/2}\sigma\\
    &\times(-1)^{p\gamma}
     \exp\left(\frac{\beta}{2}\sum^{p}_{k=1}\tr\ln\left(\Id_{2\gamma/\beta}
                      -\frac{\beta u}{8p\eta\Lambda_k}\sigma^{-1}\right)\right) \ .
  \end{split} 
\label{asymptestimate1}
\end{align}
Expanding the logarithm to leading order in $1/p$ shows that we have
to choose
\begin{eqnarray}
\eta = 
\frac{1}{p} \sum_{k=1}^p \frac{1}{\Lambda_k} =
\frac{1}{p} \tr\Lambda^{-1} \ ,
\label{etadef}
\end{eqnarray}
to fix the microscopic scale, provided  $\eta$ converges to a non zero constant for $p\rightarrow \infty$. The large--$p$ limit of the above expression is then
\begin{align}
(-1)^{p\gamma}\Det^{\gamma}\Lambda
    \Det^{p\beta/2}\sigma\exp\left(-\frac{\beta^2
        u}{16}\tr\sigma^{-1}\right) \ .
\label{asymptestimate}
\end{align}
The factor $(-1)^{\gamma p}\Det^\gamma \Lambda$ cancels
with the $p$ dependent part of $K_{p\times \bar n}$.  Combining
$\det^{p\beta/2}\sigma$ with Eq.~(\ref{InghamSiegelDistResult}) and
performing a series of integrations by parts leads to 
\begin{align}
  \begin{split}
    \mathcal{E}^{(\beta)}(u) &= K_\gamma \exp\left(-\frac{\beta
        u}{8}\right) \int\leb{\sigma} \exp\left(-\tr \sigma \right) \\
    &\times f_{\beta,2-\beta}(\sigma)\exp\left(-\frac{\beta^2 u}{16}\tr\sigma^{-1}\right) \ ,
  \end{split}
\label{emicro}
\end{align}
which can be evaluated either by using $\delta$ functions or by proper contour integration.  Remarkably, the very same matrix model results on the microscopic scale in the case that all empirical eigenvalues are equal to $1/\eta$, as
seen from Eq.~\eqref{finalESUSY},
\begin{align}
  \begin{split}
E_p^{(\beta)}\left(\frac{u}{4p\eta}\right)\Bigg|_{\Lambda=\Id_p/\eta}
 &= K_{\gamma}
    \exp\left(-\frac{\beta u}{8}\right) \int\leb{\sigma}
    \exp\left(-\tr \sigma \right) \\ &\times
    f_{\beta,\bar n}(\sigma)\Det^{p\beta/2} \left(\frac{\beta u}{8p}\Id_{ 2\gamma/\beta}-\sigma\right)\ ,
  \end{split} 
\label{estimatefinalESUSY} 
\end{align} 
where $\eta$ dropped out on the right--hand side. Thus, we have effectively traced back the problem to the uncorrelated case considered in Refs.~\cite{Edelman1991,GuhrWettigWilke,Forrester1993709,DamgaardNishigaki,KatzavPerez}. Using the scaling Eq.~(\ref{umicro}), with $\eta$ given by Eq.~(\ref{etadef}), the results coincide with the formulae of Ref.~\cite{DamgaardNishigaki}, 
\begin{align}
  \begin{split}
    \mathcal{E}^{(\beta)}(u) &= \exp\left(- \frac{\beta u}{8}\right)
    \Det^{\beta/2}\left[\tilde q_{ij}~L^{(0)}_{ij}(u)\right] ,
  \label{asympbeta2final}
  \end{split}
\end{align}
where
$L^{(l)}_{ij}(u) = \sqrt{u/4}^{i+j -\kappa'} \text{I}_{\kappa'+\delta_{i-l,0} -i-j}\left(\sqrt{u}\right)$ and $\text{I}_\nu$ is the modified Bessel function of order $\nu$. We also have defined $\kappa'=2(\gamma+1)/\beta$ and $\tilde q_{ij}=(j-i)$ for $\beta=1$, $\tilde q_{ij}=(-1)^{i+1}$ for $\beta=2$. The distribution of the smallest eigenvalue on the microscopic scale is then given by
  \begin{align}
    \begin{split}
      \wp^{(\beta)}_{\footnotesize\text{min}}(u) &= \frac{\beta}{8}\mathcal{E}^{(\beta)}(u) -\frac{\beta}{8\sqrt{u}}\exp\left(-\frac{\beta u}{8}\right) \\
        &\times\frac{\sum\limits_{l=1}^{2\gamma/\beta}\det\left[\tilde  q_{ij}~L^{(l)}_{ij}(u)\right]}{\Det^{1-\beta/2}\left[\tilde  q_{ij}~L^{(0)}_{ij}(u)\right]}\ .
    \end{split} 
 \label{pminasymptoticbeta2}
  \end{align}
To illustrate this new universality, we carry out numerical simulations for $ \wp^{(2)}_{\footnotesize\text{min}}(u)$ shown in Fig.~\ref{fig}. We generate  30,000 Hermitian Wishart correlation matrices with $p=200$ and $n=202$, for a non-trivial empirical correlation matrix $C$ as indicated in Fig.~\ref{fig}.

In conclusion, we have calculated the gap probability and the
distribution of the smallest eigenvalue for the complex and the real
correlated Gaussian Wishart ensemble. By numerical evaluation of
our results we demonstrated that they are easy to use in applications.
We also found a new universality on the microscopic scale.
On the conceptual level, our most important result is the discovery of
the duality between the $W$ and the $\overline{W}$ matrix models.
Actually, there are infinitely many dualities, as there is full
freedom in choosing that dimension of the matrices which corresponds
to the number of time steps. In turn, each of these models has a dual
model in superspace with, in general, different bosonic and fermionic
dimensions. Of course, here we chose the simplest duality that led to
a model in a superspace which collapses to an ordinary space, because
the bosonic dimension is zero. A presentation with further results and more mathematical
details will be given elsewhere \cite{WirtzGuhrII}.

\begin{acknowledgments}
  We thank Rudi Sch\"afer for fruitful discussions on applications of
  our results. We are grateful to Mario Kieburg and Santosh Kumar for
  many useful comments. We acknowledge support from the Deutsche
  Forschungsgemeinschaft, Sonderforschungsbereich Transregio 12.
\end{acknowledgments}

\bibliography{ref.bib}{}

\end{document}